\documentclass[prl,twocolumn,superscriptaddress,floatfix]{revtex4}
\pdfoutput=1
\usepackage{latexsym}
\usepackage{amsthm}
\usepackage{amssymb}
\usepackage{amsmath}
\usepackage[
pdfauthor={Jacob R. West, Bryan H. Fong, Daniel A. Lidar}, 
pdftitle={Near-optimal dynamical decoupling of a qubit},
pdfsubject={quantum dynamical decoupling}, 
pdfkeywords={quantum, dynamical decoupling, DD, QDD}]{hyperref}
\usepackage{graphicx}

\DeclareMathOperator{\bigO}{\mathcal{O}}

\DeclareMathOperator{\CDD}{CDD}
\DeclareMathOperator{\PDD}{PDD}

\DeclareMathOperator{\UDD}{UDD}
\DeclareMathOperator{\QDD}{QDD}

\newcommand{\bra}[1]{\left\langle#1\right\vert}
\newcommand{\ket}[1]{\left\vert#1\right\rangle}

\newcommand{\norm}[1]{\left\Vert #1\right\Vert}

\newcommand{\set}[1]{\left\{#1\right\}}

\begin{document}
\title{Near-optimal dynamical decoupling of a qubit} 
\author{Jacob R. \surname{West}} 
\author{Bryan H. \surname{Fong}} 
\affiliation{HRL Laboratories, LLC., 3011 Malibu Canyon Rd., Malibu,
  California 90265, USA}  

\author{Daniel A. Lidar} 
\affiliation{Departments of Chemistry, Electrical Engineering, and
  Physics, Center for Quantum Information \& Technology, University of
  Southern California, Los Angeles, California 90089, USA}

\date{July 7, 2009}

\begin{abstract}
  We present a near-optimal quantum dynamical decoupling scheme that
  eliminates general decoherence of a qubit to order $n$ using
  $\bigO(n^2)$ pulses, an exponential decrease in pulses over all
  previous decoupling methods.  Numerical simulations of a qubit
  coupled to a spin bath demonstrate the superior performance of the
  new pulse sequences.
\end{abstract}

\maketitle

Quantum information processing requires the faithful manipulation and
preservation of quantum states.  In the course of a quantum
computation, uncontrolled coupling between a quantum system and its
environment (or bath) may cause the system state to decohere and
deviate from its desired evolution, potentially resulting in a
computational error.  Here we present a dynamical decoupling
(DD)\cite{Viola:99} scheme designed to mitigate this effect.  DD
combats this decoherence by suppressing the system-bath interaction
through stroboscopic pulsing of the system, an idea which can be
traced to the spin-echo effect, with a long tradition in NMR
\cite{Slichter:book}.  Our new DD scheme is near-optimal, and provides
an exponential improvement over all previously known DD protocols. It
suppresses arbitrary coupling between a single qubit and its
environment to $n^{\mathrm{th}}$ order in a perturbative expansion of
the total qubit-bath propagator, using $\bigO(n^{2})$ pulses.

The most general interaction between a qubit and a bath can be modeled
via a Hamiltonian of the form $H=I\otimes B_{I}+X\otimes
B_{X}+Y\otimes B_{Y}+Z\otimes B_{Z}$, where $X$, $Y$ and $Z$ are the
Pauli matrices, $I$ denotes the identity matrix, and $B_{\alpha }$ are
arbitrary Hermitian bath operators.  The term $B_{I}$ is the internal
bath Hamiltonian.  Our scheme builds upon two recent insights.  The
first is due to Uhrig \cite{Uhrig:07}, who -- for single-qubit
decoherence consisting of pure dephasing errors ($H_{Z}=I\otimes
B_{0}+Z\otimes B_{Z}$) -- found a scheme (``Uhrig DD'' -- UDD) which
prescribes a sequence of $X$ pulses ($\pi $-rotations around the
$\hat{x}$ axis) at times
\begin{equation}\label{uhrigtimes}
  t_{j} = T\sin^{2}\left(\frac{j\pi}{2n+2}\right),  
\end{equation}%
where $T$ is the total evolution time and $j=1,2,\ldots,n$, if $n$ is
even, and $j=1,2,\ldots,n+1$, if $n$ is odd. These times characterize
a filter function \cite{Uys:09,Biercuk:09} that removes the qubit-bath
coupling to $n^\mathrm{th}$ order in a perturbative expansion of the
total system-bath propagator.  UDD is provably optimal in that it
achieves the minimum number of pulse intervals, $n+1$, required to
accomplish this removal \cite{Yang:08,comment}.  For single-qubit
decoherence consisting of pure spin-flip errors ($H_{X}=I\otimes
B_{0}+X\otimes B_{X}$), UDD is still optimal, simply by substituting
$Z$ pulses ($\pi $-pulses around the $\hat{z}$ axis) for $X$ pulses.
However, for systems subject to general errors as prescribed by $H$,
DD schemes incorporating both $X$ and $Z$ pulses are required.  One
such scheme, which provides our second source of insight, is
concatenated DD (CDD) \cite{KhodjastehLidar:05}: the CDD sequence is
capable of eliminating arbitrary qubit-bath coupling to order $n$ at a
cost of $\bigO(4^{n})$ pulses \cite{KhodjastehLidar:07}. CDD works by
recursively nesting a pulse sequence found in Ref.~\cite{Viola:99},
capable of canceling arbitrary decoherence to first order. Uhrig
recently introduced a hybrid scheme (CUDD) which reduces the pulse
count to $\bigO(n2^{n})$ for exact order $n$ cancellation \cite{CUDD}.
By appropriately concatenating the UDD sequences for $H_{Z}$ and
$H_{X}$ we show here how arbitrary decoherence due to $H$ can be
exactly canceled to order $n$ using only $(n+1)^2$ pulse intervals.  A
numerical search we conducted found that this is very nearly optimal
for small $n$, differing from the optimal solutions by no more than
two pulses.

\textit{Near-optimal pulse sequence construction}.---The goal of our
construction is to integrate an $X$-type $\UDD_n$ sequence, which
suppresses pure dephasing error to order $n$, with a $Z$-type $\UDD_n$
sequence, which suppresses longitudinal relaxation to order $n$, so
that the resulting sequence removes arbitrary error to order $n$.
Since the total time $T$ in Eq.~\eqref{uhrigtimes} is arbitrary, what
matters for error cancellation is not the precise pulse times $t_j$,
but rather the \emph{relative sizes} of the pulse intervals $\tau_j
\equiv (t_j - t_{j-1})$, for $j = 1,2,\dots,n+1$.  Thus, the most
relevant quantities are normalized pulse intervals,
\begin{equation}\label{intervals}
  s_j \equiv \frac{t_j-t_{j-1}}{t_1-t_0} = \sin\left(\frac{(2j-1)
      \pi}{2n+2}\right) \csc\left(\frac{\pi}{2n+2}\right),
\end{equation}
again with $j = 1,2,\dots, n+1$.  Here we chose to normalize with
respect to the shortest pulse interval $(t_1-t_0) = t_1$, so that
$\tau_j = s_j t_1$, which has the important consequence that the
(normalized) total time grows with $n$, as pulse intervals are added
to address higher order error.  This fixing of the minimum pulse
interval $t_1$ corresponds to imposing a finite bandwidth constraint.
Since any physical implementation will not be able to shrink the pulse
intervals arbitrarily, but will be limited by the fastest pulsing
technology available, the change in perspective from fixed total time
to fixed minimum interval is appropriate.  The total normalized time
of a $\UDD_n$ sequence is then given by,
\begin{equation}\label{Tn}
  S_n \equiv \sum_{j=1}^{n+1} s_j = \frac{t_{n+1}}{t_1} =
  \csc^2\left(\frac{\pi}{2n+2}\right),
\end{equation}
so that the total physical time is $T = S_n t_1$.  

Let $U(\tau)$ denote the joint unitary free evolution of a qubit and
its bath for a time $\tau$, subject to the Hamiltonian $H$.  A
$Z$-type $\UDD_n$ sequence then takes the form,
\begin{equation}\label{zudd}
  Z_n(\tau) \equiv Z^{n}\, U(s_{n+1}\tau)\, Z\, U(s_n\tau) \cdots Z\,
  U(s_2\tau)\, Z\, U(s_1\tau).
\end{equation}
Note that a final $Z$ pulse is required for $n$ odd.  Define the
$X$-type $\UDD_n$ sequence $X_n(\tau)$ similarly.  Yang and Liu showed
in \cite{Yang:08} that the entire UDD sequence $X_n(t_1)$ can be
expressed as the propagator $\exp\left(-i S_n t_1 (I\otimes
  B_I^{\prime } + X\otimes B_X^{\prime }) + \bigO((\lambda S_n
  t_1)^{n+1})\right)$ (in units of $\hbar = 1$), where $\lambda \sim
\norm{H}$, provided $\lambda S_n t_1$ is sufficiently small to ensure
convergence of the time-perturbative expansion.  The important point
here is that the resulting effective Hamiltonian $H^{\prime } =
I\otimes B_I^{\prime } + X\otimes B_X^{\prime }$ is a purely
spin-flip, time-independent, coupling.  The correction term
$\bigO((\lambda S_n t_1)^{n+1})$ potentially contains all couplings
with a complicated time-dependence, but is suppressed to order $n$.
Moreover, subject to the convergence condition, we are free to choose
the minimum interval arbitrarily without impacting the validity of the
proof in \cite{Yang:08}.  This is the key to correctly integrating the
$X$-type and $Z$-type $\UDD_n$ sequences.  The desired error
cancellation properties only require that the normalized pulse
intervals $s_j$ have the specified structure.  The precise physical
timing is inconsequential.  Therefore, to integrate $X_n$ and $Z_n$
sequences properly, without breaking the delicate pulse timing
structure required for error cancellation, we must \emph{scale} the
pulse intervals of the inner sequences uniformly with respect to the
outer pulse sequence structure.  Hence, if each $U(s_j\tau)$ in
Eq.~\eqref{zudd} is replaced by the time-scaled DD sequence
$X_n(s_j\tau)$, then the outer $Z_n$ sequence suppresses the purely
spin-flip coupling $H^{\prime }$ remaining after each $X_n(s_j\tau)$
sequence, producing general decoherence suppression to order $n$ with
only $(n+1)^2$ pulse intervals.  So the combined, near-optimal
``quadratic DD'' (QDD) sequence, takes the form,
\begin{equation}\label{qdd}
  \QDD_n(\tau) \equiv Z^n\, X_n(s_{n+1}\tau)\,
  Z\, X_n(s_n\tau) \cdots Z\, X_n(s_1\tau),
\end{equation}
abbreviated by the notation $\QDD_n(\tau) = Z_n(X_n(\tau))$.  Notice
how the relative scales of the pulse intervals are preserved.  In each
of the inner $X_n(s_j \tau)$ sequences, the ratio between successive
intervals remains $(s_{k+1} s_j \tau)/(s_k s_j \tau) = s_{k+1}/s_k$,
and for the outer $Z_n$ sequence the ratio is $(S_n s_{j+1} \tau)/(S_n
s_j \tau) = s_{j+1}/s_j$, thereby ensuring the error cancellation
properties of each sequence are left intact.  Of course, an equivalent
$\QDD_n$ sequence may be constructed as $X_n(Z_n(\tau))$.  Moreover,
though the inner DD sequences must have an equal number of intervals,
they need not be the same length as the outer sequence, but can
instead be adjusted to more efficiently address the dominant sources
of error in any particular implementation.  This way, $\QDD_{m,n} =
Z_m(X_n(\tau))$ is the more general construction, where the inner
sequences suppress one type of error to order $n$, while the outer
sequence suppresses the remaining error to order $m$.  As the simplest
explicit example, $\QDD_1 = Z X_1(s_2\tau) Z X_1(s_1\tau) = Z\, (X\,
U(s_2^2 \tau)\, X\, U(s_1 s_2\tau))\, Z\, (X\, U(s_2 s_1 \tau)\, X\,
U(s_1^2 \tau)) = Y\, U(\tau)\, X\, U(\tau)\, Y\, U(\tau)\, X\,
U(\tau)$, which we recognize as the so-called ``universal decoupler''
sequence found in \cite{Viola:99} and used as the basis of the CDD
sequence in \cite{KhodjastehLidar:05}.

It is worth noting that the construction of $\QDD_n$ affords a nice
visualization.  For $\sigma \in \set{I,X,Y,Z}$, define $\sigma(s_j)
\equiv \sigma\, U(s_j \tau)\, \sigma$, then consider,
\begin{center}
\begin{tabular}[c]{c c c c c | c}
  $Y(s_{n+1}^2)$ & $Z(s_n s_{n+1})$ & $\cdots$ &
  $Y(s_2 s_{n+1})$ & $Z(s_1 s_{n+1})$ & $Z(s_{n+1})$ \\[1mm] 
  $X(s_{n+1} s_n)$ & $I(s_n^2)$ & $\cdots$ & $X(s_2 s_n)$
  & $I(s_1 s_n)$ & $I(s_n)$ \\[1mm]
  & & $\vdots$ & & & $\vdots$ \\[1mm]
  $Y(s_{n+1} s_2)$ & $Z(s_n s_2)$ & $\cdots$ & $Y(s_2^2)$
  & $Z(s_1 s_2)$ & $Z(s_2)$ \\[1mm]
  $X(s_{n+1} s_1)$ & $I(s_n s_1)$ & $\cdots$ &
  $X(s_2 s_1)$ & $I(s_1^2)$ & $I(s_1)$ \\[1mm]
  \hline 
  $X(s_{n+1})$ & $I(s_n)$ & $\cdots$ & $X(s_2)$ & $I(s_1)$
  & $\ket{\psi}$
\end{tabular}
\end{center}
with the final $\QDD_n$ pulse sequence formed by reading off rows of
the inner square from top to bottom, or columns from left to right.
In other words, our construction may be succinctly described as an
outer product between $X$-type and $Z$-type UDD sequences.

Also, note that the $\QDD_n$ sequence requires a total physical time
of $S_n^2 \tau$, and just as in the original proof \cite{Yang:08} of
order $n$ error suppression for the $Z_n$ sequence, the condition for
convergence of the perturbative expansion still remains, namely that
$\lambda S_n^2\tau$ is sufficiently small.  An immediate consequence
of this constraint is that there exists a maximal order of error $n$
that can be suppressed for a given $\tau$, beyond which the error
cancellation properties of the sequence begin to break down.
Conversely, if one hopes to suppress error to some fixed order $n$,
then this implies a maximum pulse rate $r = 1/\tau$ which must be
attained.  Specifically, a sufficient condition for convergence is
$\lambda S_n^2\tau < 1$, hence $r > \lambda\csc^4(\pi/(2n+2))$.

Finally, we translate these results back into a precise physical
timing for the individual pulses.  If the total time for the sequence
is $T$, then the outer $Z_n$ sequence requires that $Z$ pulses occur
at the original Uhrig times prescribed in Eq.~(\ref{uhrigtimes}), while
the inner $X_n$ sequences require $X$ pulses executed at the times,
$t_{j,k} = \tau_j \sin^2\left(\frac{k\pi}{2n+2}\right) + t_{j-1}$,
where $j,k= 1,2,\ldots n$ if $n$ is even and $j,k=1,2,\dots n+1$ if
$n$ is odd.  When $n$ is odd, $X$ and $Z$ pulses coincide at times
$t_j$, in which case $Y = ZX$ pulses are used.

\textit{Numerical results}.---We now present numerical results that
illustrate the efficiency of our new DD pulse sequences in preserving
arbitrary initial quantum states.  In the full-state quantum memory
simulations that follow, the system and bath are initialized together
as a (generally nonseparable) random pure state $\ket{\psi}$.  The
results presented involve a single system qubit coupled to four bath
qubits.  While this is clearly an unrealistically small bath, larger
simulations we performed indicate that the relevant error suppression
properties of our DD sequences are qualitatively unaffected by bath
size.  Similarly idealized are the DD pulses themselves, which we take
to be infinitely strong zero-width pulses, consistent with the
analysis in \cite{Yang:08}.  The evolution of the coupled system and
bath in our simulations is governed by the generic Hamiltonian
$H$,
with additional parameters $\set{J,\beta}$ included to control the
coupling and bath strengths, respectively:
\begin{equation}\label{jbham}
H = \beta\left(I\otimes B_I\right) + J\left(X\otimes B_X + Y\otimes
  B_Y + Z\otimes B_Z\right),
\end{equation}
where each bath operator is given by $B_{\alpha} = \sum_{i\not=
  j}\sum_{k,l} r_{kl}^\alpha \left(\sigma_i^k\otimes
  \sigma_j^l\right)$, with $\alpha,k,l \in \set{I,X,Y,Z}$, $i,j$
indexing the bath qubits, and randomly chosen coefficients
$r_{kl}^\alpha \in [0,1]$.  Note that these bath operators include all
$1$- and $2$-body terms, so that the system-bath Hamiltonian includes
$2$- and $3$-body terms.  The numerical simulations are run with more
than $100$ digits of precision and results are averaged over $10$
random realizations, with each instance randomly generating new bath
operators and a new initial state.  The vertical axes in each of these
plots quantifies the DD sequence performance as
$\log_{10}(\mathrm{D})$, where $\mathrm{D}$ is the standard trace-norm
distance $\frac{1}{2}\norm{\rho(T) - \rho(0)}_1$ between the evolved
system state $\rho(T) = \mathrm{tr}_B (\ket{\phi}\bra{\phi})$, with
$\ket{\phi} = \QDD_n(\tau)\ket{\psi}$, and the initial system state
$\rho(0) = \mathrm{tr}_B (\ket{\psi}\bra{\psi})$.  Here
$\mathrm{tr}_B$ is the partial trace over the bath.  This distance
measure bounds the usual Uhlman fidelity $F$ from above and below,
$1-D \leq F \leq \sqrt{1-D^2}$, and is itself bound by the norm of a
perturbative expansion of the propagator \cite{LZK:08}.  Indeed, in
our simulations, $D$ always goes like the norm of leading order term,
as one might expect.  Error bars are included at every point
indicating the maximum deviation from the average distance measured.
\begin{figure}[htp]
\centering
\includegraphics[width=87mm]{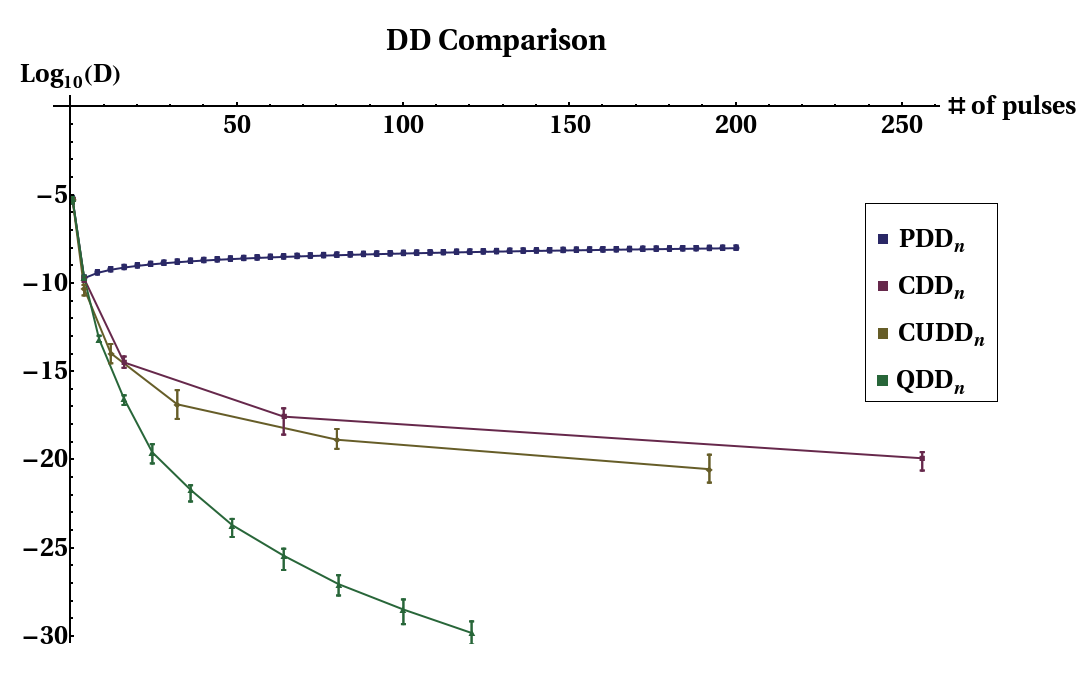}
\caption{Performance comparison of various DD schemes vs. number of
  pulses required.  $\QDD_n$ dramatically outperforms all known
  alternatives in both state preservation and number of pulses.}
\label{fig:ddcomp}
\end{figure}

Figure \ref{fig:ddcomp} shows the performance of various DD schemes
plotted against the number of pulses each requires.  $\PDD_n$ is the
$n$-times repeated universal decoupling sequence found in
\cite{Viola:99}, requiring $4n$ pulses (this sequence is capable only
of first order decoupling); $\CDD_n$ is the $n$-times concatenated
universal decoupling sequence, requiring $4^n$ pulses
\cite{KhodjastehLidar:05}; $\mathrm{CUDD}_n$ is concatenated Uhrig DD,
combining an $n^{\mathrm{th}}$ order $X$-type Uhrig sequence with $n$
concatenation levels of $Z$ pulses, requiring $n2^n$ pulses
\cite{CUDD}.  The coupling parameters $J$ and $\beta$ are fixed so
that $J\tau = \beta\tau = 10^{-6}$, given a shortest pulse interval
$\tau$.  Recall that since the shortest pulse interval is held
constant, the total evolution time grows with increasing $n$.
Specifically, for $\QDD_n$ the total evolution time is $S_n^2 \tau$.
The evolution times of the other DD sequences also increase with $n$,
but at different rates depending on how the number of required pulse
intervals scales with $n$.  Notice the two most visible consequences
of increasing total time in Figure \ref{fig:ddcomp}: (1) at $n=0$ the
evolution time is only $S_0^2 \tau =\tau$, so the dominant
contribution to the overall fidelity is $J\tau = 10^{-6}$, which
explains why the undecoupled evolution still achieves excellent
agreement with the initial state in this plot; and (2) as the total
time increases so does decoherence, explaining why the performance of
$\PDD_n$ actually degrades with large $n$.

\begin{figure}[htp]
\centering
\includegraphics[width=87mm]{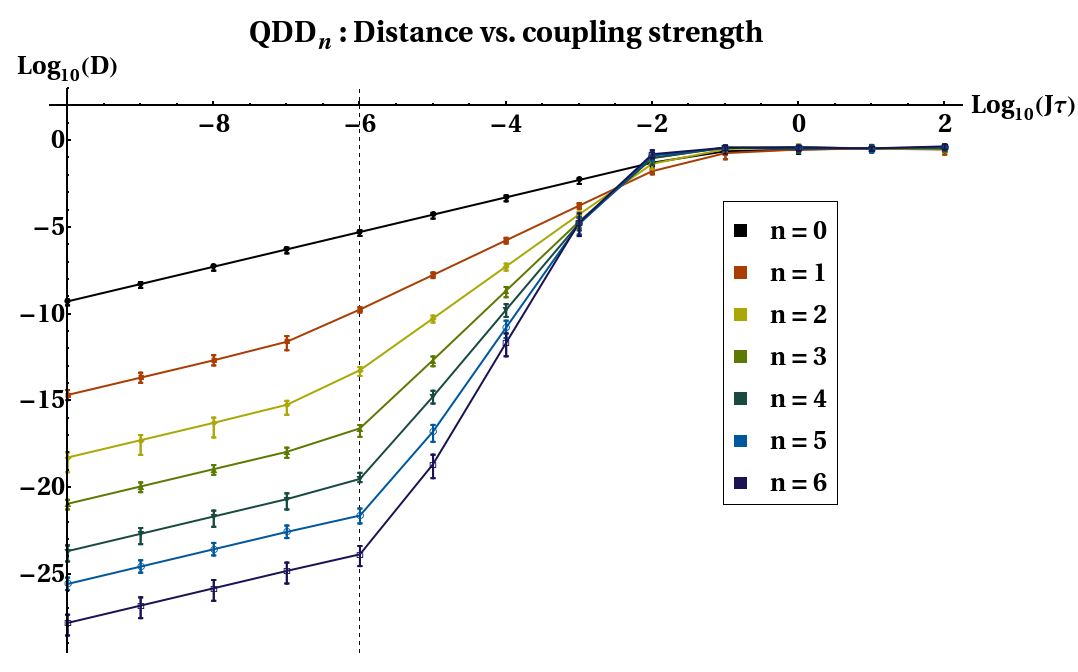}
\caption{$\QDD_n$ performance dependence on coupling strength $J$,
  with fixed $\protect\beta\protect\tau = 10^{-6}$.}
\label{fig:qddj}
\end{figure}

\begin{figure}[htp]
\centering
\includegraphics[width=87mm]{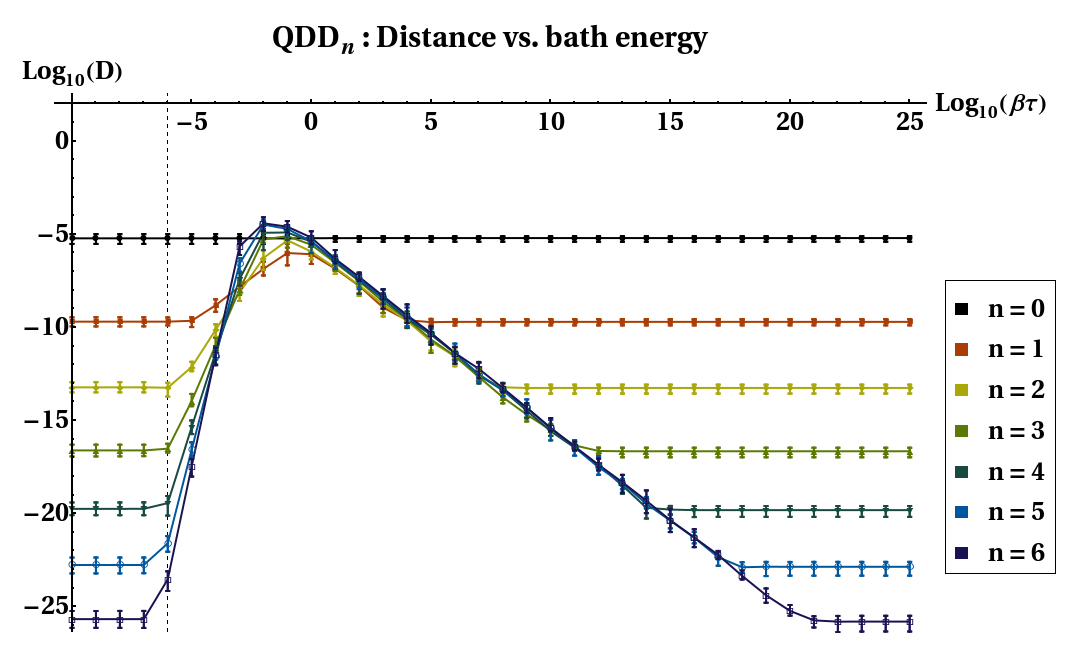}
\caption{$\QDD_n$ performance dependence on bath strength, with fixed
  $J\protect\tau = 10^{-6}$.}
\label{fig:qddb}
\end{figure}

Figure \ref{fig:qddj} shows $\QDD_n$ performance as a function of $J$,
the system-bath coupling strength.
As $J S_n^2\tau$ approaches 1, the qubit decoheres so rapidly that DD
has essentially no effect, and the distance between the initial and
final states approaches its maximum of 1.  Again, the $n=0$ line
represents undecoupled free evolution, for which $J S_n^2\tau = 1$
when $J\tau = 1$, corresponding to the point in this plot where the
$\log_{10}(\mathrm{D})$ becomes zero and subsequently stays there.
State preservation improves as $n$ increases and another order of
error is suppressed, with the effect magnified as $J\tau$ decreases,
though increasing total time counteracts the overall performance gain.
This is evidenced in how, at each fixed $J\tau$, the magnitude of the
improvement decreases as $n$ increases, or in other words, in how the
gap between lines narrows as $n$ increases.  The dotted line in both
Figures \ref{fig:qddj},\ref{fig:qddb} indicates when $J\tau =
\beta\tau$.  Across this transition point, the slope increases as the
leading order error term changes from $\beta$ to $J$ dominated in
Figure \ref{fig:qddj}, and vice versa in Figure \ref{fig:qddb}.
Indeed, when $J < \beta$, the leading order term contributes to
distance as $\frac{(n J\beta^n)(S_n^2\tau)^{n+1}}{(n+1)!}$, which
gives a slope of $1$ as a function of $J$ for every $n$ in Figure
\ref{fig:qddj}.  On the other hand, when $J > \beta$, the dominant
contribution to performance becomes $\frac{(J
  S_n^2\tau)^{n+1}}{(n+1)!}$, explaining the observed slopes of $n+1$
in this plot.

Figure \ref{fig:qddb} shows the performance dependence of $\QDD_n$
vs. $\beta$, the parameter which quantifies the pure bath energy
scale.  
We again see the $J\tau = \beta\tau$ transition in this figure (the
dotted line), as well as the improvement with increasing $n$.  The
obviously significant feature of this plot however is the deviation,
and then eventual returning, of the evolved state relative to the
initial state as $\beta\tau$ sweeps through the regimes: $(1)\,
\beta\tau < J\tau < S_n^{-2}$, $(2)\, J\tau < \beta\tau < S_n^{-2}$,
and $(3)\, \beta\tau > S_n^{-2}$.  This behavior can be understood
mathematically by considering an interaction picture with respect to
the pure bath dynamics, governed by $H_B = \beta(I\otimes B_I)$, then
expanding the other bath operators in the $B_I$ eigenbasis.  Using a
Dyson series expansion with $J S_n^2\tau < 1$, we find that when
$\beta\tau < J\tau$, the distance goes like $\frac{(J
  S_n^2\tau)^{n+1}}{(n+1)!}$, as in the previous plot, independent of
$\beta$; when $J\tau < \beta\tau < S_n^{-2}$, the distance goes like
$\frac{(n J\beta^n)(S_n^2\tau)^{n+1}}{(n+1)!}$, giving a slope of $n$
as a function of $\beta$; and when $\beta\tau > S_n^{-2}$, the
distance goes like $\left[(C\frac{J}{\beta} + 1) \frac{(J
    S_n^2\tau)^{n+1}}{(n+1)!}\right]$ ($C$ is a constant), which gives
a slope of $-1$ until $\beta$ is sufficiently large as to eliminate
the contribution of the first term, leaving a leading order term that
goes again like the small $\beta$ regime.

\textit{Conclusions}.--- We have presented a DD pulse sequence
construction that guarantees simultaneous cancellation of both
transverse dephasing and longitudinal relaxation to order $n$ using
only $(n+1)^2$ pulse intervals, or $\bigO(n^2)$ pulses, an exponential
improvement over all previous DD methods.  The constructed sequences
are near-optimal, in that an exhaustive numerical search we performed
produced optimal pulse sequences of lengths $7$, $14$, and $23$, for
$n = 2, 3$, and $4$, respectively, i.e., $(n+1)^2-2$ pulses.  In view
of this, we believe our construction is very nearly optimal, in
addition to having the significant conceptual benefit of a simple
algorithmic description. Moreover, it is worth noting that the filter
function may alternatively be optimized for specific bath noise
spectra (Locally Optimized Dynamical Decoupling,
LODD)\cite{Biercuk:09} or over a range of bath frequencies (Optimized Noise Filtration 
through Dynamical Decoupling, OFDD)\cite{Uys:09}, resulting
in pulse timings different from Eq.~\eqref{uhrigtimes} and decoupling
performance superior to both CPMG \cite{Slichter:book} and UDD for
high-frequency baths \cite{Biercuk:09,Uys:09}.  Our construction can
be easily applied to LODD and OFDD, but for ease of explanation we
restricted our discussion to the pulse timing in
Eq.~\eqref{uhrigtimes}.  Finally, of particular importance will be
generalizing this work to incorporate finite width pulses and,
ultimately, computation \cite{Khodjasteh:09}.  We look forward to
experimental tests of these sequences.

\textit{Acknowledgments}.---The authors thank Mark Gyure and Richard
Ross for their feedback on early drafts of this paper.  Sponsored by
United States Department of Defense.  The views and conclusions
contained in this document are those of the authors and should not be
interpreted as representing the official policies, either expressly or
implied, of the U.S. Government.  Approved for public release,
distribution unlimited.


\begin{thebibliography}{10}
\bibitem{comment}
Optimality follows from the fact that there are $n$ algebraic constraints
satisfied by $n+1$ pulse intervals for $\bigO(n)$ decoupling.
\end{thebibliography}
\end{document}